\documentstyle[prl,aps,psfig]{revtex}
\draft

\newcommand{\be}{\begin{equation}}
\newcommand{\ee}{\end{equation}}
\newcommand{\bary}{\begin{eqnarray}}
\newcommand{\eary}{\end{eqnarray}}
\newcommand{\ga}{\Gamma_{\perp}}
\newcommand{\gaa}{\Gamma_{\parallel}}
\newcommand{\ggt}{\Gamma_T}
\newcommand{\dmev}{\Delta m^2/eV^2}
\newcommand{\dm}{\Delta m^2}
\newcommand{\sint}{sin^2 2\theta}

\newcommand{\lsim}{\mathrel{\rlap{\raisebox{.3ex}{$<$}}
    \raisebox{-.6ex}{$\sim$}}}
\begin{document}
\title{ Effect of random magnetic field on
active-sterile\\
neutrino conversion in the supernova core
}
\author{
Sarira Sahu$^{*}$ and Vishnu M. Bannur}
\address{ 
Institute For Plasma Research,
Bhat, Gandhinagar-382 428, India}
\maketitle
\begin{abstract}
\noindent  
The active-sterile neutrino conversion is studied for neutrino propagating
in the axial potential generated by magnetised electron plasma in the supernova 
medium. We consider the effect of random magnetic field $B_{rms}$ on the 
average neutrino conversion probability. We obtained the constraint on 
$\dm$ and $\sint$ for different strength of the random magnetic fields,
by considering the positive definiteness of the average neutrino 
conversion probability inside the supernova core. Our calculation shows that,
$B_{rms}\lsim 0.63\times 10^{14}$ Gauss is preferable so that small values
of $\dm$ can not be excluded.

\end{abstract}
\section{Introduction}

  The COBE measurement of the cosmic microwave background temperature
anisotropies on large scale\cite{cobe}, hints for the existence of a hot 
dark matter (HDM) component of 30\% of the total mass density\cite{hdm}. 
Also the recent results of solar neutrino, atmospheric neutrino and the LSND
result, hints for the neutrino oscillation. All these can not be explained
with the three generation neutrino flavor mixing scheme. Thus the 
speculation is that, there is a fourth generation low mass  sterile neutrino
$\nu_s$ which will mix with the standard model neutrinos. This neutrino 
has to be sterile with respect to electroweak interactions so that, 
it can not be detected directly through experimental searches. This
postulate of a fourth generation light-sterile neutrino can 
explain simultaneously the dark matter, solar neutrino and atmospheric 
problems\cite{valle,anjan,peltoniemi,akhmedov}. The
most stringent constraints for the neutrino mass matrix including a sterile
neutrino species are obtained from the nucleosynthesis bound on the maximum
number of extra neutrino species that can reach thermal equilibrium before
nucleosynthesis and change the primordially produced helium 
abundance\cite{walker}.

Neutrino propagation in the magnetised medium has interesting consequences 
in the astrophysical and cosmological scenarios. Large scale  magnetic fields 
in the early universe hot plasma and in the core of the supernova can effect
the neutrino conversion\cite{zeldo}. It has been shown that random magnetic 
fields can
strongly influence neutrino conversion rates and this could have important
implications, especially in the case of conversion involving a light sterile
neutrino\cite{pas1,pas2}.
The effect of active-sterile  neutrino conversions in a supernova has 
also been discussed, both in the case where no magnetic field is present, as
well as in the presence of random magnetic field\cite{pas2} as large as 
$10^{16}$ Gauss.

The dispersion relation for neutrino propagating in a magnetised medium is
different from the vacuum one. In the magnetised medium neutrino acquires an
axial potential which is proportional to the scalar product of the neutrino
momentum and the magnetic field vector ($\bf k.B$). The effect of axial
potential on neutrino propagation in media with regular and/or random magnetic
fields are considered in the literature\cite{pas2,sahu1,jose,semikoz}. 
For random magnetic fields case 
Semikoz and Valle \cite{semikoz} have show that, the neutrino conversion is 
aperiodic in nature. The effect of axial potential on active-sterile 
neutrino conversion
has also been considered in supernova and sun\cite{pas2,smirnov}. 
It was shown by Kusenko and
Segre recently that, the polarisation effects of the supernova medium lead to
the explanation of the birth velocity of pulsars\cite{kusenko}.

   In this paper we have considered the effect of axial potential on neutrino
propagation in the supernova medium in the presence of random magnetic fields.
We calculate the average neutrino
conversion probability ${\cal P}$, for the process 
$\nu_e\rightarrow\nu_s$ using two different approximations (method-I and II). 
method-I shows
that, for both $\gaa > \omega_f$ and $\gaa < \omega_f$ 
($\gaa$ is the longitudinal damping parameter) one can obtain
finite conversion probability for $\nu_e\rightarrow\nu_s$. On the other hand
for the positive definiteness of the conversion probability in method-II,
only $\gaa < \omega_f$ is possible. Using this positive definiteness condition
of the average conversion probability inside a supernova core, we have
showed that $B_{rms}\lsim 0.63\times 10^{14}$ Gauss
is preferred, because large values of $B_{rms}$ exclude
small values of $\dm$, which are not supported by present experiments.

The paper is organised as follows: in Sec. II we calculate the average neutrino
conversion probability using two different methods in the presence of a 
randomly fluctuating magnetic field. The comparison of both the methods
are undertaken in Sec. III and necessary condition for the positive definiteness
of the average conversion
probability for the process $\nu_e\rightarrow\nu_s$ is obtained in the second
method. Using the necessary condition for active-sterile neutrino 
conversion inside a supernova core, we put 
constraint on the parameters $\dm$ and $\sint$ for different 
values of the random magnetic field in Sec. IV and a short conclusion is drawn 
in Sec. V.

\section{Neutrino propagation in the magnetised medium}

\subsection{method - I}
The evolution equation for a system of two neutrinos $\nu_a$ and $\nu_b$, 
where $\nu_a$ is the active one and $\nu_b$ is active/sterile one is given by
\be
i{d\over dt}{\pmatrix {\nu_a\cr \nu_b\cr}}
 = {\pmatrix {H_{aa}(t) & H_{ab}(t)\cr H_{ba}(t) & H_{bb}(t)\cr}}
 {\pmatrix {\nu_a\cr \nu_b\cr}},
\label{evu}
\ee
where the quantity $H$  is in general the potential for the neutrino
in the medium which we will discuss soon.
Let us define the functions $R= Re(\langle\nu_a^*\nu_b\rangle)$ and
$I=Im(\langle \nu_a^*\nu_b\rangle)$. Then using these in Eq.(\ref{evu})
we obtain
\be
{\dot R}(t)=-H_d(t) I(t), ~~~~~
{\dot I}(t)=H_{ab}(t) \left ( 2 P(t)-1\right ) + H_d R(t),
\label{reim}
\ee
and
\be
{\dot P}(t)=-2 H_{ab}(t) I(t)
\label{prob}
\ee
where  the function $P(t)$ is the neutrino conversion probability
$P_{\nu_a\rightarrow\nu_b}(t)$ and $H_d=H_{aa}(t)-H_{bb}(t)$
and dot on the top corresponds to derivative with respect to $t$.
Using the Eq.(\ref{reim}) in Eq.(\ref{prob}) we can write
\bary
{\dot P}(t)&=& -2 H_{ab}(t)\int_0^t {\dot I}(t_1)dt_1\nonumber\\
&=&-2 H_{ab}(t)\int_0^t
\left [ H_{ab}(t_1) (2 P(t_1)-1) + H_d(t_1) R(t_1)\right ] dt_1.
\label{pdot}
\eary
Let us consider the neutrino propagation in the medium in the presence 
of a magnetic field. Then we have
\be
H_d(t)=V-\Delta cos 2\theta+V_{axial} ~~~and~~~ H_{ab}(t)=\mu B_{\perp}(t).
\ee
The quantity $V$ is the difference of neutrino vector potential for
$\nu_a\rightarrow\nu_b$, 
$\Delta=(m_2^2-m_1^2)/2E=\Delta m^2/2E$ and
$E$ is the neutrino energy and $\theta$ is the neutrino mixing angle.
The axial vector potential $V_{axial}=\mu_{eff}{\bf k.B}(t)/k$ is generated
by the mean axial vector current of charged leptons in an external magnetic 
field\cite{sahu1,jose,semikoz,espo}.
For fluctuation in the magnetic field we can write
$B(t)=B_0+B'(t)$, where $B_0$ is the constant background field 
and $B'(t)$ is the random fluctuation over it. Then we have 
\bary
H_d(t) &=& H_d(0) + H'_d(t)\nonumber\\ 
&=& (V-\Delta+\mu_{eff}B_{\parallel 0}) +\mu_{eff}B'_{\parallel}(t),
\eary
and
\bary
H_{ab}(t) &=& H_{ab}(0) + H'_{ab}(t)\nonumber\\
&=& \mu B_{\perp 0}+\mu B'_{\perp}(t).
\eary
For the neutrino conversion length
greater than the domain size i.e. $l_{conv}>>L_0$
(where $l_{conv}\sim 1/{{\cal P}\Gamma_W}$ and $\Gamma_W$ is the 
weak interaction rate), a neutrino will cross
many magnetic field domains before it flips its helicity.
Thus the neutrino will experience an average field before it flips its
helicity. So one can average the propagation equation
(\ref{pdot}) over the random magnetic field distribution\cite{pas1,enq1}. 
The magnetic field in different domains is randomly oriented with respect to
the neutrino propagation direction. So the neutrino conversion probability
depends on the root mean square ({\it rms}) value of the random magnetic field.
With the use of the delta correlation
for uncorrelated magnetic field domains of size $L_0$, the average of the
random magnetic field is\cite{semikoz,enq1,enq2,sahu,bala},
\be
\langle B_{\parallel}(t)\rangle =
\langle B_{\perp}(t)\rangle =
\langle B_{\parallel}(t)B_{\perp}(t)\rangle =0,
\label{cor1}
\ee
\be
\langle B_{i\parallel}(t)B_{j\parallel}(t_1)\rangle
=\langle B^2_{\parallel}\rangle \delta_{ij}L_0\delta(t-t_1),
\ee
\label{cor2}
and
\be
\langle B_{i\perp}(t)B_{j\perp}(t_1)\rangle 
=\langle B^2_{\perp}\rangle \delta_{ij}L_0\delta(t-t_1).
\label{cor3}
\ee
The {\it rms} value of the averaged magnetic field is given as
$B_{rms} = \sqrt{\langle B^2\rangle }$.
Let us assume, 
\begin{itemize}
\item
$H_{ab}(t)$ is sufficiently close to $H_{ab}(t_2)$ so that
\be
\int_0^t\int_0^{t_1} H_{ab}(t)
\frac{H_d(t_1)H_d(t_2)}{H_{ab}(t_2)}{\dot P}(t) dt_2 dt_1
\simeq
\int_0^t\int_0^{t_1}
H_d(t_1)H_d(t_2){\dot P}(t) dt_2 dt_1,
\label{appx1}
\ee
and
\item
$P(t)$ has no correlation with $H_d(t)$ and $H_{ab}(t)$. 
\end{itemize}
Thus averaging both sides of Eq.(\ref{pdot}) and using the
above magnetic field correlations we obtain
\be
{\dot{\cal P}}(t)=
-2 H^2_{ab}(0) \langle \int_0^t (2P(t_1)-1)dt_1 \rangle
-\ga \langle 2P(t)-1)\rangle 
-H^2_d(0)\langle \int_0^t P(t_1)dt_1\rangle - 2\gaa\langle \int_0^t
{\dot P}(t_1) dt_1\rangle.
\label{avpd}
\ee
where $\langle P\rangle = {\cal P}$ and 
the quantities $\ga$ and $\gaa$ are the longitudinal and transverse 
damping parameters given by
\be
\ga=\frac{4}{3}\mu^2 \langle B^2\rangle L_0, ~~~ and ~~~
\gaa = \frac{1}{6}\mu_{eff}^2 \langle B^2\rangle L_0.
\label{damp}
\ee
Differentiating once again to Eq.(\ref{avpd}) with respect to $t$ we obtain
the following second order differential equation 
\be
{\ddot{\cal P}}(t) + 2\ggt {\dot{\cal P}}(t) 
+\omega_s^2{\cal P}(t)-2 H^2_{ab}(0)=0,
\label{master1}
\ee
where the quantity
\be
\omega_s^2=4H^2_{ab}(0) +  H^2_d(0),
\ee
is square of the spin  
rotation frequency of the neutrino in the medium, $\ggt=\ga+\gaa$ and it
satisfies the boundary conditions
${\cal P}(0)=0$ and ${\dot{\cal P}}(0)=\ga$.
The solution to this differential equation is given by,
\be
{\cal P}(t)=
\frac{2H^2_{ab}(0)}{\omega_s^2}
\left (
1-e^{-\ggt t}
\left \{
\frac{\sinh[\sqrt{\ggt^2-\omega_s^2}~t]}{\sqrt{\ggt^2-\omega_s^2}}
[\gaa-\frac{\ga(\omega_s^2-2H^2_{ab}(0))}{2H^2_{ab}(0)}]
+\cosh[\sqrt{\ggt^2-\omega_s^2}~t]
\right \}
\right ).
\label{pt1}
\ee
The Eq.(\ref{pt1}) is the general solution for the average conversion 
probability having random fluctuation in both transverse and longitudinal 
mode of the magnetic field\cite{semikoz}. 
For no random fluctuation in the magnetic field ($\ga=\gaa=0$) we get back
the standard MSW type solution for the neutrino conversion probability
\be
P(t)=\frac{4 H^2_{ab}(0)}{\omega_s^2} \sin^2(\frac{\omega_s t}{2}).
\label{mswm}
\ee
Let us consider the propagation of a system of active (doublet) 
and light sterile (singlet) neutrinos ($\nu_e\rightarrow\nu_s$),
with masses $m_1$ and $m_2$, mixing angle $\theta$,
and no transition magnetic moments, in the presence of  a magnetised plasma.
Then the Hamiltonian in the evaluation equation Eq.(\ref{evu}) will be
\be
{\pmatrix {V-\Delta \cos2\theta+\mu_{eff}{\bf k.B}/{k} 
& \Delta\sin2\theta/2\cr \Delta \sin2\theta/2 & 0\cr}},
\label{mix}
\ee
where $\Delta=(m_2^2-m_1^2)/2E$.
For active-sterile neutrino conversion the resultant vector potential 
experienced by $\nu_e$ is given by
\be
V=\sqrt{2} G_F n_e (3Y_e + 4 Y_{\nu_e} -1),
\ee
where $G_F$ is the Fermi coupling constant, $n_e$ is the electron density in the
medium and $Y_e$ and $Y_{\nu_e}$ are the electron and $\nu_e$
abundances respectively in the medium. 
For neutrino propagating along the $z$ axis the $V_{axial}$ is
\be
V_{axial}=\mu_{eff} B_z\frac{k_z}{k}.
\ee
The quantity $\mu_{eff}$ for $\nu_e\rightarrow\nu_s$ is given by
\be
\mu_{eff}=\frac{e G_F P_F}{\sqrt{2}~ 2\pi^2},
\ee
and $P_F$ is the Fermi momentum of electron.
As we are considering the neutrino magnetic moment/transition magnetic moment
to be zero, the perpendicular component of the damping term $\ga$ will vanish.
Thus putting $\ga=0$ in Eq.(\ref{pt1}), 
the average conversion probability for $\nu_a\rightarrow\nu_s$ will be
\be
{\cal P}(t)=\frac{\Delta^2\sin^2 2\theta}{2\omega_f^2}
\left (
1-e^{-\gaa t}
\left \{
\gaa\frac{\sinh[\sqrt{\gaa^2-\omega_f^2}~t]}{\sqrt{\gaa^2-\omega_f^2}}
+\cosh[\sqrt{\gaa^2-\omega_f^2}~t]
\right \}
\right ),
\label{mmsw}
\ee
where
\be
\omega_f^2=(H^2_d(0)+\Delta^2 sin^22\theta)
\ee
is square of the flavor 
conversion frequency of the neutrino in the medium.
In a supernova medium with a strong random magnetic field satisfying the
condition
$\gaa\gg \omega_f$ the average conversion probability in Eq.(\ref{mmsw})
will be approximately 
\be
{\cal P}(t) \simeq\frac{\Delta^2\sin^2 2\theta}{2 \omega_f^2}
\left ( 1- e^{-\omega_f^2 t/{2\gaa}}\right ),
\label{strong}
\ee
and for weak field 
limit i.e.  $\gaa \ll \omega_f$ it will be
\be
{\cal P}(t) \simeq\frac{\Delta^2\sin^2 2\theta}{2 \omega_f^2}
\left ( 1- e^{-\gaa t}\cos\omega_f t \right ).
\label{weak}
\ee
These are shown previously by Semikoz and Valle in ref\cite{semikoz}.
Using the strong field limit for the active-sterile neutrino conversion
in a supernova, the limit on $\dm$ and $\sint$ are obtained from the
supernova cooling\cite{pas2} and by thermalisation of the sterile neutrinos
in the early universe hot plasma\cite{semikoz}.

\subsection{method - II}

  In the previous section we have derived the master equation for the 
average conversion probability of neutrino in the medium with a magnetic field.
But in that calculation we have neglected the correlation of $P(t)$ with
$H_d(t)$ and $H_{ab}(t)$. In the previous section we have only taken the 
average of the equation for
${\dot P}(t)$ and not for ${\dot R}(t)$ and ${\dot I}(t)$ to derive the
master equation for the conversion probability. 
For the neutrino conversion length 
greater than the domain size ($l_{conv}>>L_0$), a neutrino will cross
many magnetic field domains before it flips its helicity.
Thus the neutrino will experience an average field before it flips its
helicity. So one can average the propagation equations 
(\ref{reim}) and (\ref{prob}) over the random
magnetic field distribution\cite{pas1,enq2}. Let us define the average of the
functions $\langle P(t)\rangle ={\cal P}(t)$, $\langle R(t)\rangle 
={\cal R}(t)$ and 
$\langle I(t)\rangle ={\cal I}(t)$. 
Because of the averaging the average probability
${\cal P}(t)$ will only depend on the even powers of the magnetic 
field correlation. 
Using the average functions in Eqs.(\ref{cor1}) to (\ref{cor3}) we
obtain
\bary
{\dot {\cal I}}(t) &=& \langle H_{ab}(t) (2P(t)-1)\rangle
+ \langle H_{d}(t)R(t)\rangle \nonumber\\
&=& H_{ab}(0) \langle (2P(t)-1)\rangle + \langle {\tilde H_{ab}}(2P(t)-1)\rangle
\nonumber\\
& & +H_{aa}(0)\langle R(t)\rangle + \langle {\tilde H_{aa}}(t)R(t)\rangle,
\label{idot}
\eary
and
\be
{\dot {\cal R}}(t) = -H_{aa}(0)\langle I(t)\rangle - 
\langle {\tilde H_{aa}}(t) I(t)\rangle 
\label{rdot}
\ee
respectively. Using the delta correlations for the magnetic fields as shown
in Eqs.(\ref{cor1}) to (\ref{cor3}) we obtain
\be
\langle H'_d(t)R(t)\rangle \simeq -2\gaa {\cal I}(t);
~~~
\langle H'_{ab}(t)P(t)\rangle \simeq -\ga {\cal I}(t),
\ee
and
\be
\langle H'_{ab}(t)I(t)\rangle \simeq \frac{\ga}{2} (2 {\cal P}(t) -1).
\ee
Putting these values in Eqs.(\ref{idot}) and (\ref{rdot}) we obtain
\be
{\dot {\cal I}}(t) =
H_{d}(0) {\cal R}(t) + H_{ab}(0) (2 {\cal P}(t) - 1)
-2 (\ga + \gaa) {\cal I}(t).
\label{ci}
\ee
and
\be
{\dot{\cal R}}(t) = - H_{d}(0) {\cal I}(t) - 2 \gaa {\cal R}(t).
\label{cr}
\ee
Using these equations in Eq.(\ref{prob}) we obtain
\be
{\dot{\cal P}}(t) = -2H_{ab}(0){\cal I}(t)-{\ga}\left (2{\cal P}(t)-1
\right ).
\label{pdot2}
\ee
For convenience let us define
\be
{\cal I}(t) = e^{-2(\gaa+\ga)t}{\cal I}_{1}(t),
\label{i1}
\ee     
and
\be
{\cal R}(t) = e^{-2\gaa t}{\cal R}_{1}(t),
\label{r1}
\ee
Using Eqs.(\ref{i1}) and (\ref{r1}) in Eqs.(\ref{ci}) to (\ref{pdot2}) we obtain
\be
{\dot {\cal I}_1}(t) = 
H_{d}(0) e^{2\ga t}{\cal R}_1(t) + H_{ab}(0) e^{2(\ga+\gaa) t}
(2 {\cal P}(t) - 1),
\label{nidt}
\ee
\be
{\dot{\cal R}}_1(t) = - H_{d}(0) e^{-2\ga t}{\cal I}_1(t), 
\ee
and
\be
{\dot{\cal P}}(t) = -2H_{ab}(0)e^{-2(\ga+\gaa)t}{\cal I}_1(t)
-{\ga}\left (2{\cal P}(t)-1\right ).
\label{npt}
\ee
Differentiating Eq.(\ref{npt}) twice with respect to $t$ and putting the 
value of ${\dot {\cal I}_1}(t)$ from Eq.(\ref{nidt}) we obtain the  
master equation for the average conversion probability as,
\bary
{\stackrel{\ldots}{\cal P}}(t) &+&
4\left (\ga+\gaa\right )  {\ddot{\cal P}}(t) 
+ 4\left ( 3\ga\gaa+
\ga^2+\gaa^2
+\frac{\omega_s^2}{4}\right ){\dot {\cal P}}(t)
\nonumber\\
&+&8\left (\ga^2\gaa
+\ga\gaa^2
+H_{ab}^2(0)\gaa
+{H_{d}^2(0)\ga\over 4}\right ){\cal P}(t)
\nonumber\\
&-&4\left (\ga^2\gaa
+\ga\gaa^2
+H_{ab}^2(0)\gaa+{H_{d}^2(0)\ga\over4}\right )=0,
\label{master2}
\eary
with the boundary conditions ${\cal P}(0)=0,~~{\dot {\cal P}}(0)=\ga$ and 
${\ddot {\cal P}}(0)=2H_{ab}^2(0)-2\ga^2$.
By switching off the damping terms in Eq.(\ref{master2}), we will give
get back the same solution for P(t) as shown in Eq.(\ref{mswm}).
The solution to the above third order differential equation is
given in Eq.(25) of the ref\cite{sahu} and it is of the form
\be
{\cal P}(t)=\frac{1}{2} + y(t).
\ee
Henceforth for further reference we will refer the Eq.(25) of ref\cite{sahu}. 
The solution obtained is very complicated
and it is difficult to conclude any thing from the solution. On the other 
hand the interesting part of the solution is that, the positive definiteness
of the average conversion probability ($0\le {\cal P}\le 1$)  requires the 
following condition to satisfy
\be
\omega^2_s ~>~{4\over 3} (\ga^2 + \gaa^2 - \ga\gaa),
\label{cond}
\ee
irrespective of the form of neutrino potential and the magnetic field,
which essentially shows that neutrino crosses many domains leads to this
requirement.
Considering the magnetic field in the early universe hot plasma and the
core of the newly born neutron stars to be purely random in nature, we 
studied the conversion 
$\nu_{eL}\rightarrow\nu_{eR}$ in these medium in a previous paper\cite{sahu}. 

\section {comparison of both the methods}

Let us consider the active sterile neutrino conversion in the magnetised
medium, with neutrino mixing and $\mu=0$ ($\ga=0$) as shown in (\ref{mix}). 
Only the fluctuation in 
the parallel component of the magnetic field
will contribute. So for this case the  condition in Eq.(\ref{cond}) will be
modified to
\be
\omega^2_f ~>~\frac{4\gaa^2}{3}.
\label{cond2}
\ee
This implies  $\omega_f >\gaa$.
The average conversion probability in Eq.(\ref{strong}) for strong random 
magnetic field limit is derived with the approximation $\gaa\gg \omega_f$.
On the other hand we obtain from Eq.(\ref{cond2}), completely the opposite one.
Also the later condition further shows that, what ever may be the strength 
of the random component of the magnetic field $\gaa$, it should always
satisfy the condition in Eq.(\ref{cond2}). 
Apart from that, this
condition comes automatically from the positive definiteness of the
probability. 
This opposite situation arises because of the averaging 
procedure. In the first case we only take the average of the Eq.(\ref{prob})
and use the functions $I(t)$ and $R(t)$ in that. On the other hand in the
second method, we average the equations for $P(t)$, $I(t)$ and $R(t)$ 
separately, which are shown explicitly. In the first method it is assumed that
$H_{ab}(t)$ is sufficiently close to $H_{ab}(t_2)$ so that in 
Eq.(\ref{appx1}), they cancel. But in general this is not true for 
arbitrary $t$ and $t_2$.
Such terms are not there in the second method 
to derive the average probability equation. 

Comparison of both the master equations Eq.(\ref{master1}) and 
Eq.(\ref{master2}) shows that, in the first there is no mixing between the
longitudinal $\gaa$ and transverse $\ga$ damping terms. Also the damping 
terms do not mix with the background terms. On the other hand, there is
mixing between the damping and background terms, and the probability equation
becomes more complicated.

\section{results and discussions}

Let us estimate the range of $\dm$ and $\sint$ from the
the above inequality in Eq.(\ref{cond2}) in a supernova environment.
Inside the supernova the vector potential for the process
$\nu_e\rightarrow\nu_s$ is
\be
V\simeq 4\times 10^{-6}\rho_{14} (3 Y_e+4 Y_{\nu_e}-1) MeV
\simeq 4.48~ eV,
\ee
where $Y_e\simeq 0.3$, $Y_{\nu_e}\simeq 0.06$ and the quantity 
$\rho_{14}\simeq 8$ is the density in the supernova core
in units of $10^{14} gm/cm^3$. 
Now let us assume that the random fluctuation in the magnetic field is much
large compared to the constant background part i.e. 
$\mu B_{\parallel 0}\ll \gaa$ as considered in ref\cite{pas2}. 
The flavor precession frequency for $\nu_e\rightarrow\nu_s$ is given by
\be
\omega_f=\sqrt{(V - \frac{\Delta m^2}{2 E} \cos 2\theta)^2 + 
\left (\frac{\Delta m^2}{2 E}\right )^2 \sin^2 2\theta}.
\ee
Inside the supernova core, neutrinos have energy in the range 30 to 100 MeV.
Then considering $E\simeq 100$ MeV, we obtain
\be
\omega_f=\sqrt{(4.48 - 0.5\times 10^{-8}\frac{\Delta m^2}{eV^2} 
\cos 2\theta)^2 + 
0.25\times 10^{-16} \left (\frac{\Delta m^2}{eV^2}\right )^2 
\sin^2 2\theta}~~ eV.
\label{omegaf}
\ee
Inside the supernova core for $\nu_e\rightarrow\nu_s$, 
\be
\mu_{eff}\simeq 4.3\times 10^{-13}\mu_B\left (\frac{P_F}{MeV} \right ),
\ee
where $\mu_B$ is the Bohr magneton. The quantity $\mu_{eff}$ has the same
dimension that of the magnetic moment, but it has nothing to do with the
magnetic moment as it does not change the helicity of the particle.
The Fermi momentum of the electron inside the core is 
\be
P_F\simeq 320 (Y_e\rho_{14})^{1/3}~~ MeV \simeq 428~~ MeV.
\ee
Thomson and Dunkan have argued that very strong magnetic fields 
might be generated inside the supernova core 
due to small scale dynamo mechanism. If these fields are generated 
after core collapse, then it could be viewed as random superposition of many
small dipoles of size $L_0\sim 1$ Km. 
Then the longitudinal damping parameter is simplified to
\be
\gaa \simeq 9.6 B^2_{14}~~eV, 
\ee
where $B_{rms}$ is expressed in units of $10^{14}$ Gauss. Putting $\omega_f$ and
$\gaa$ in Eq.(\ref{cond2}), we can find the ranges of $\dm$ 
and $\sint$ for which the condition in Eq.(\ref{cond2}) is satisfied. 
These are shown in
the contour plots for different values of the magnetic fields in Figure 1. 
Figure 1 (a) shows that, a very narrow
range of $\dmev$ ($3.16\times 10^8-1.6\times 10^9$)
is excluded for $0\lsim \sint 0.63\times 10^{-3}$ for the random 
magnetic field strength $B_{rms}=0.1\times 10^{14}$ Gauss. 
By increasing the strength of the random magnetic field to $0.5\times 10^{14}$ 
and $0.63\times 10^{14}$ Gauss, we see in Figure 1 (b) and (c) that the 
width of the excluded range of $\dmev$ has increased. 
The spread of $\dmev$ towards smaller values is large compared to the 
large values. Also the large values of $\sint$ are excluded. 
In Figure 1 (b) we see that $3.5\times 10^{8}\lsim\dmev\lsim 1.6\times 10^9$
is excluded and for this the excluded range of $\sint$ is
$0\lsim \sint \lsim 0.4$ (for $B_{rms}=0.5\times 10^{14}$ Gauss).
In Figure 1 (c)
it is shown that for $B_{rms}=0.63\times 10^{14}$ Gauss, the excluded region
of $\dmev$ is $1.6\times 10^7~-~1.6\times 10^9$ and in this range all the
values of $\sint$ are excluded. It shows that, two distinct
allowed regions for $\dmev$ are there in both sides of the excluded region.
In these allowed regions all ranges of $\sint$ are allowed.  
But so far as $\dm$ value is concerned, it is interesting to consider only
the region left to the excluded curve, because it corresponds to smaller
$\dmev$ values ($\dm \lsim (keV)^2$).
Going from Figure 1 (c) to (d) (for $B_{rms}=10^{14}$ Gauss)
we observe that, the left arm of the
curve vanishes and the right arm spreads towards the higher values of 
$\dmev$ for all ranges of $\sint$. This implies that small 
values of $\dmev$ are excluded. 
By further increasing the strength of the random magnetic field, we have
shown in Figure 1 (e) that, (for $B_{rms}=10^{16}$ Gauss) all values of 
$\dmev \lsim 1.6\times 10^{13}$ are 
excluded for all ranges of $\sint$ values. This range of $\dmev$ are obviously
not at all interesting, because it corresponds to very high values of $\dmev$,
and no astrophysical, cosmological and laboratory observations favour this.
Also if we consider small
values of magnetic field ($B_{rms} < 10^{13}$ Gauss) then we found that,
the excluded region vanishes and all the parameter ranges are allowed.
Thus if we consider only the
effect of random magnetic field in a magnetised electron
plasma along the neutrino propagation direction, then magnetic field should
satisfy $B_{rms}\lsim 10^{14}$ Gauss so that small $\dmev$ ranges
should not be excluded. On the other hand for $B_{rms}< 10^{13}$ Gauss
we found that, all the parameter ranges are allowed. 
As we have shown in Figure 1 (a), only a very
narrow strip of $\dmev$ values are excluded for 
$0 \lsim \sint \lsim 0.4$ and this
narrow strip vanishes for smaller values of $B_{rms}$.
For $\dm$ and $\sint$ very small, we can see in Eq.(\ref{omegaf}) that
$\omega_f\simeq 2.12$ eV and the condition Eq.(\ref{cond2}) will give
$B_{rms} < 0.4\times 10^{14}$ Gauss. 
Thus we found that for magnetic field $B_{rms}\lsim .63\times 10^{14}$ Gauss, 
all the interesting ranges of parameters are allowed. 
Also this implies that the maximum value of $\dm$ can be in the
$(keV)^2$ range. 

\section{conclusion}

For neutrino propagating in the  magnetised plasma of the supernova, will 
experience an axial potential which is proportional to the scalar product 
of the neutrino momentum and the magnetic field. 
We derived the average conversion probability for active neutrino goes to
active or sterile one by using a method developed by Semikoz et. al, and
another by one of us where it is assumed that the magnetic field has a 
random fluctuation over the constant background. In the second method 
for neutrino having non-zero mixing and zero magnetic
moment we found the condition for the positive definiteness of the average
conversion probability for the process $\nu_e\rightarrow\nu_s$. 
Using this condition as the basis, we found the
excluded/included ranges of $\dm$ and $\sint$ for different values of the
magnetic field. Our calculation shows that
for neutrino to have maximum mass in the keV range or less, the random 
magnetic field along the neutrino propagation direction should not be very
large ($ B_{rms} \lsim 0.63\times 10^{14}$ Gauss). Because it is observed that
for large magnetic field, small values of $\dm$ are excluded.

\vfill\eject

\vspace*{2.2in}
\begin{figure}
\includegraphics{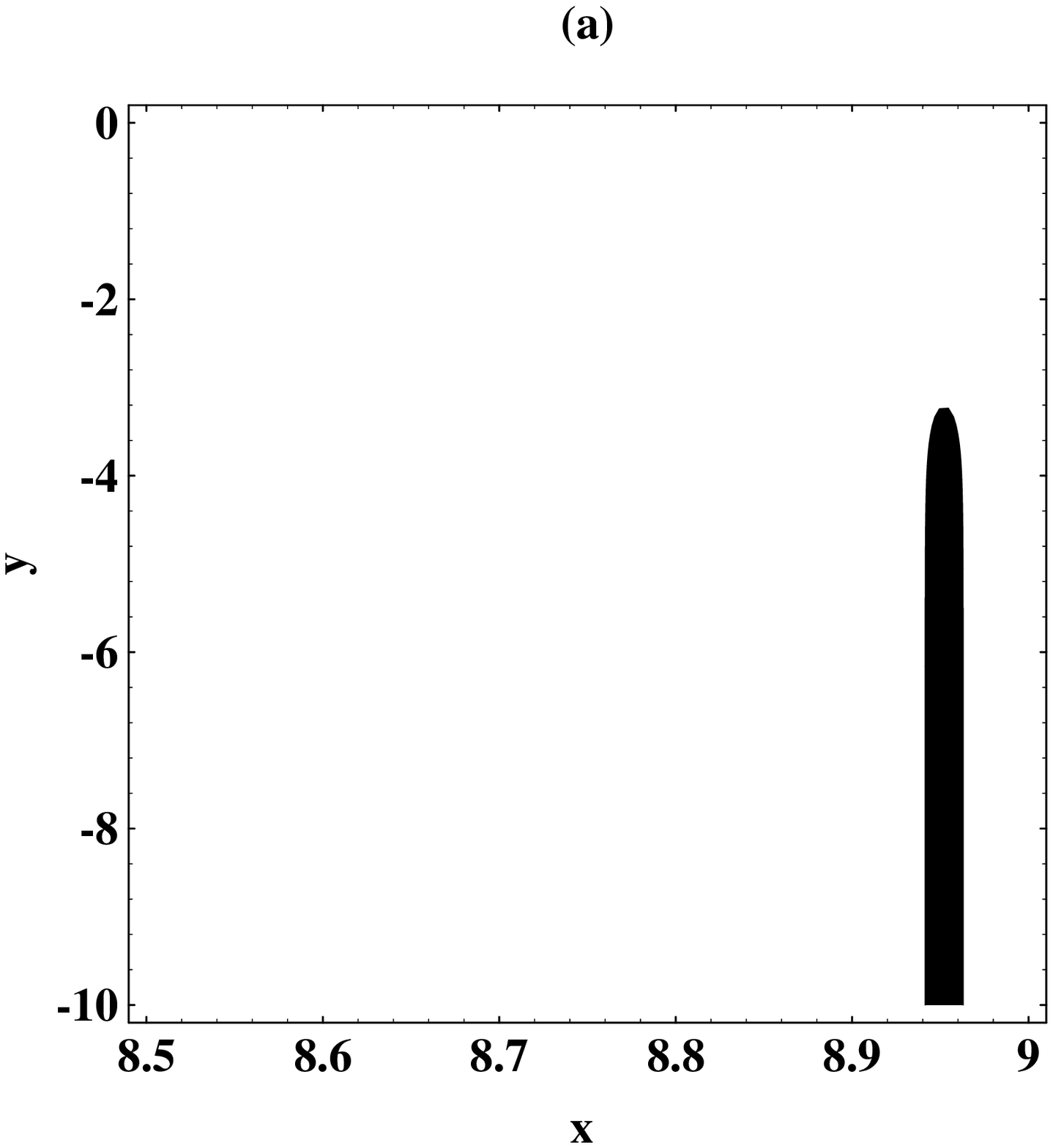}
\includegraphics{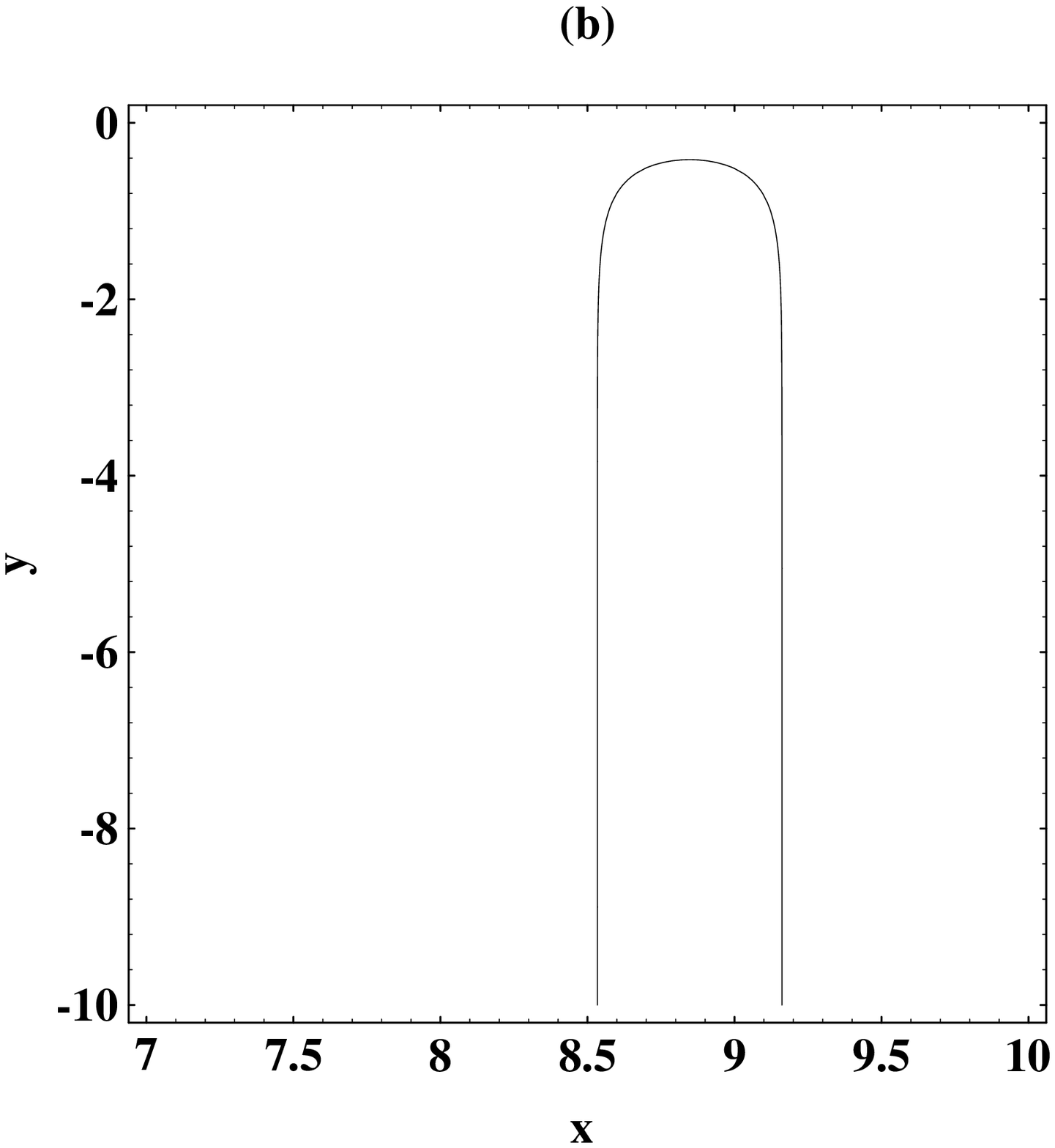}
\includegraphics{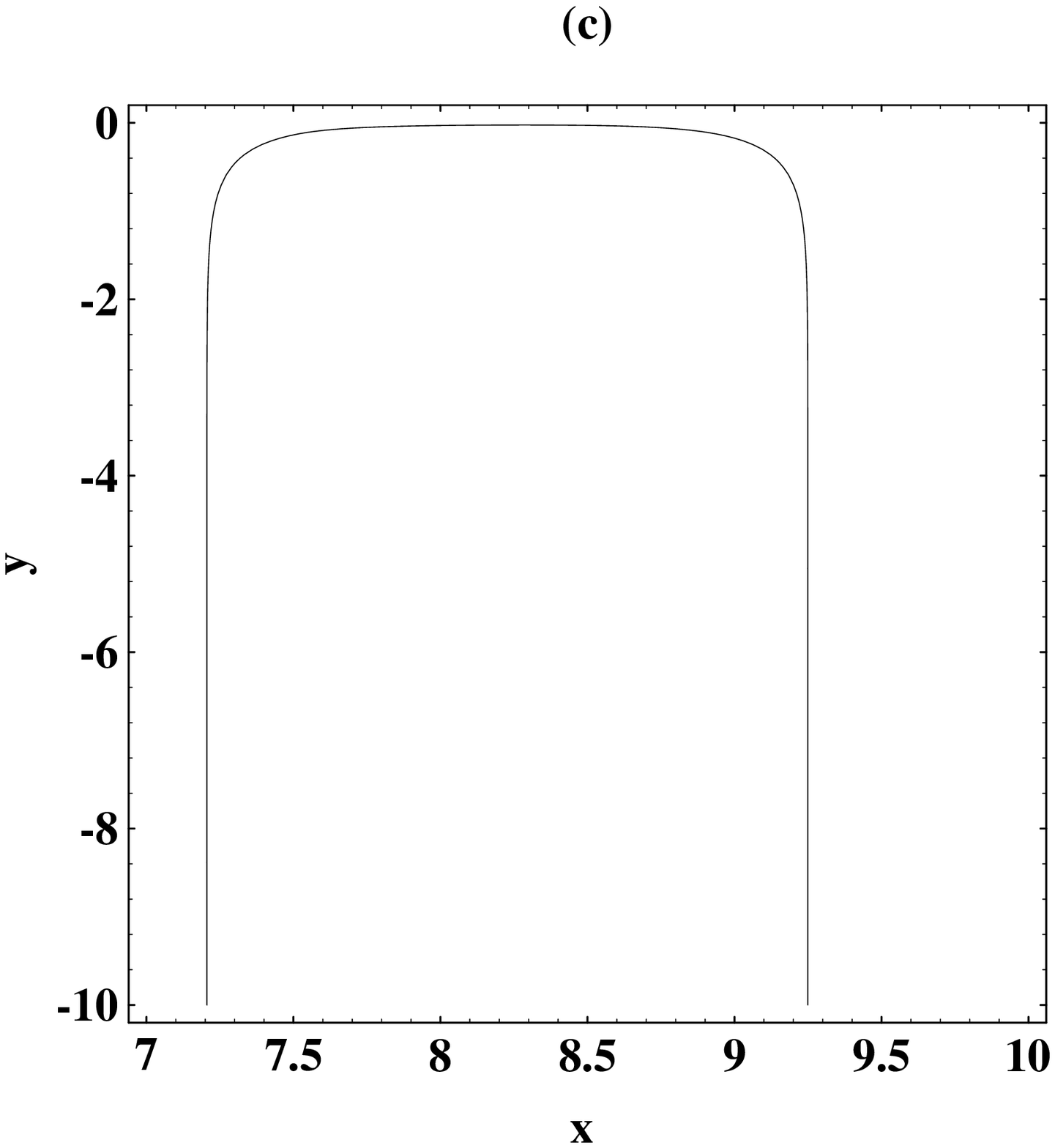}
\includegraphics{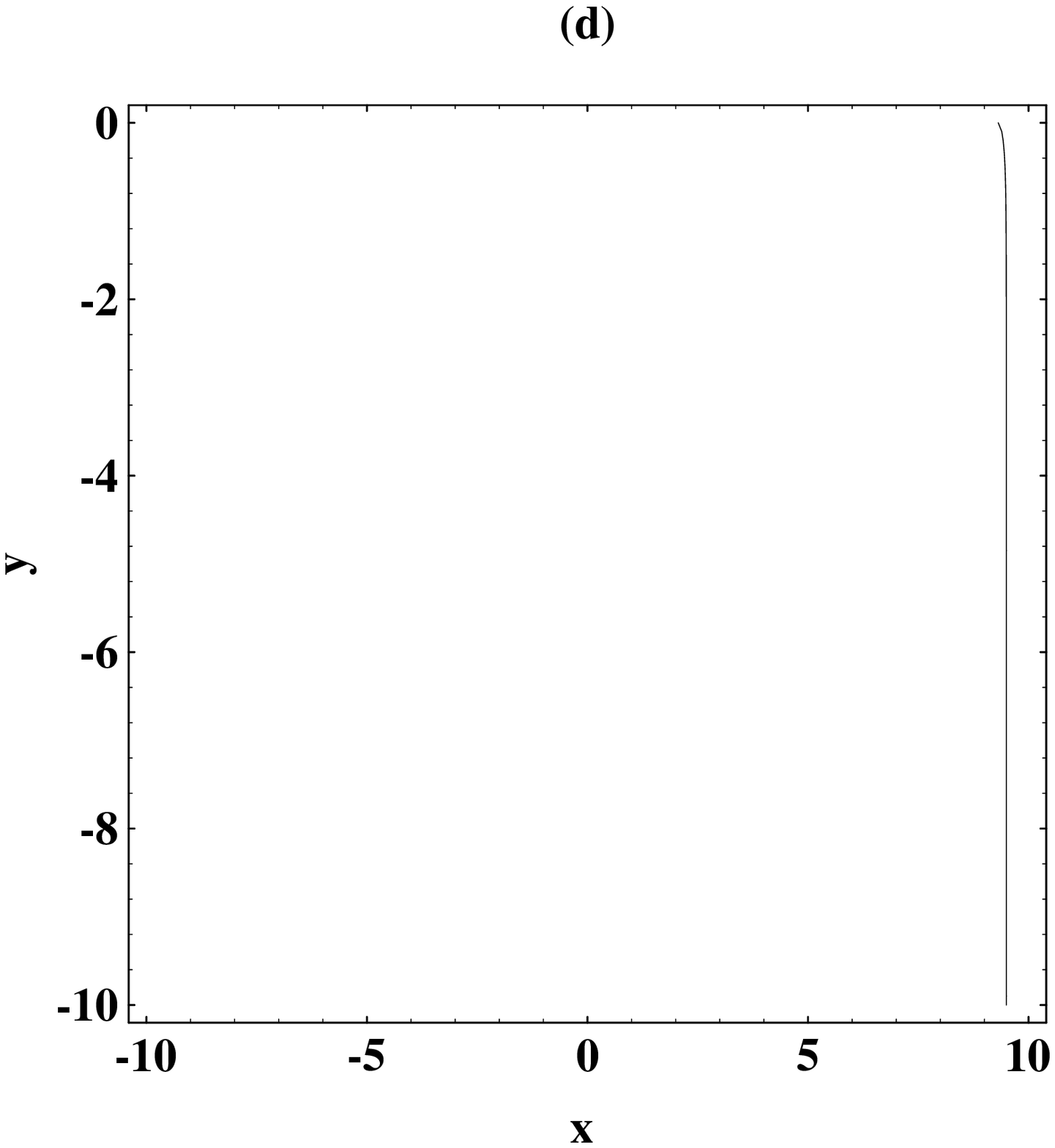}
\includegraphics{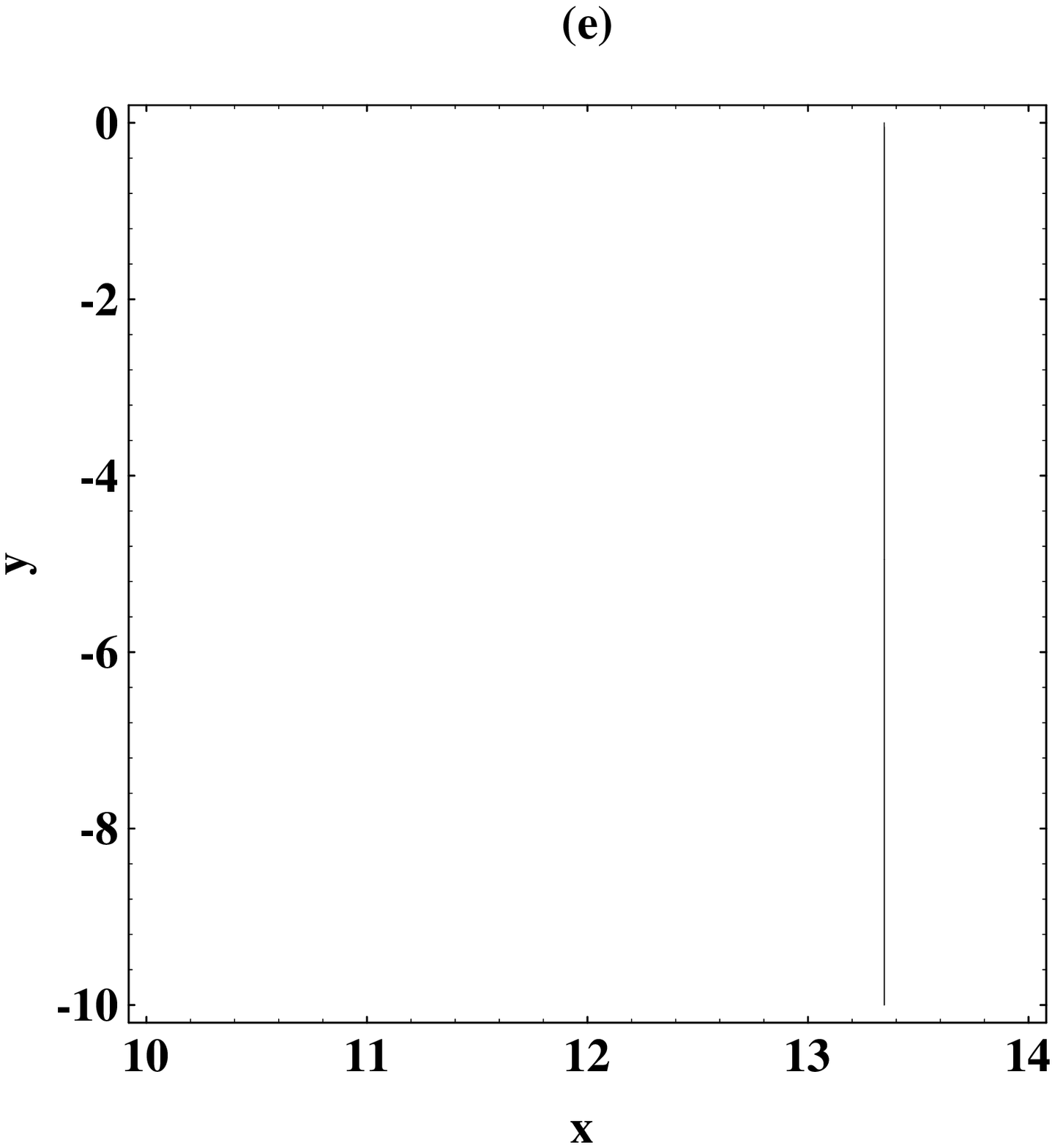}
\end{figure}
\vspace*{5.2in}
{Figure 1: We have defined ${\bf x}=log[\Delta m^2/eV^2]$
and ${\bf y}=log[sin^2 2\theta]$.(a) It is for $B_{rms}=10^{13}$ 
Gauss and the shaded
region is excluded, (b) for $B_{rms}=0.5\times 10^{14}$ Gauss and
the region inside the curve is excluded, (c) $B_{rms}=0.63\times 10^{14}$
Gauss and the region inside the curve is excluded, (d) for 
$B_{rms}=10^{14}$ Gauss and left side of the curve is excluded and (e)
$B_{rms}=10^{16}$ Gauss and the left side of the curve is excluded.
}

\end{document}